\documentclass[fleqn,12pt,twoside]{article}
\usepackage{espcrc1}

\usepackage{graphicx}
\usepackage[figuresright]{rotating}

\newcommand{\AmS}{{\protect\the\textfont2
  A\kern-.1667em\lower.5ex\hbox{M}\kern-.125emS}}

\title{Approaching the dynamics of hot nucleons in supernovae}

\author{M. Liebend\"orfer\address[CITA]{CITA, University of Toronto,
Toronto, Ontario, M5S 3H8, Canada}, U. Pen\addressmark[CITA]
and C. Thompson\addressmark[CITA]}

\begin{document}

\maketitle

\begin{abstract}
All recent numerical simulations agree that stars in the main sequence
mass range of \( 9-40 \) M\(_{\odot}\) do not produce a prompt
hydrodynamic ejection of the outer layers after core collapse and
bounce.
Rather they suggest that stellar core collapse and supernova explosion
are dynamically distinct astrophysical events, separated by an unspectacular
accretion phase of at least \(\sim 40\) ms duration. As long as the
neutrinospheres remain convectively stable, the explosion dynamics
is determined by the neutrons, protons, electrons and neutrinos in
the layer of impact-heated matter piling up on the protoneutron star.
The crucial role of neutrino transport in this regime has been
emphasized in many previous investigations.
Here, we search for efficient means to address the role of magnetic
fields and fluid instabilities in stellar core collapse and the
postbounce phase.
\end{abstract}

\section{Introduction}

The complicated trajectory through core collapse determines the state of the
cold nuclear matter deep inside the protoneutron star. At its surface
and above, the hot mantle of shock-dissociated nucleons grows by the
continued accumulation of infalling matter, heated by the impact at
the fairly stationary accretion front at a radius of order \(100\) km.
Simulations with general relativistic
neutrino transport in spherical symmetry have explored, in detail,
the important role of weak interactions between neutrinos and nuclear
matter in stratified layers of curved space-time
\cite{Bruenn_DeNisco_Mezzacappa_01,Liebendoerfer_et_al_03}. Before \( 50 \) ms
after bounce, the entropy of the shocked matter is
even higher than the entropy that would be achieved by infinitely
long exposure to the prevailing neutrino field. After that,
the neutrinospheres recede to smaller radii and produce higher luminosities.
The accretion front moves to larger radius where the
gravitational potential is weaker and the infalling matter has less
kinetic energy. Only after that time, do the conditions develop for
neutrino heating to become effective. As illustrated in Fig.~1,
incoming fluid elements experience an entropy increase toward
the equilibrium entropy set by the prevailing neutrino abundances.
If the fluid element is compressed during infall, antineutrinos
are preferentially absorbed to adjust the electron fraction in response
to the increased electron Fermi energies; if it expands, more neutrinos
are absorbed. The infalling fluid element crosses
the equilibrium entropy at the gain radius. Due to the small reaction
time scale in the cooling region, the state of the fluid element
approximately follows the local equilibrium entropy and electron
fraction thereafter until the neutrinos become trapped behind the
energy-dependent neutrinospheres. The caption of
Fig.~1 encapsulates the well-known reasons why spherically symmetric
supernova models have failed to explain the observed explosions.

\begin{figure}[htb]
\includegraphics[width=0.465\textwidth]{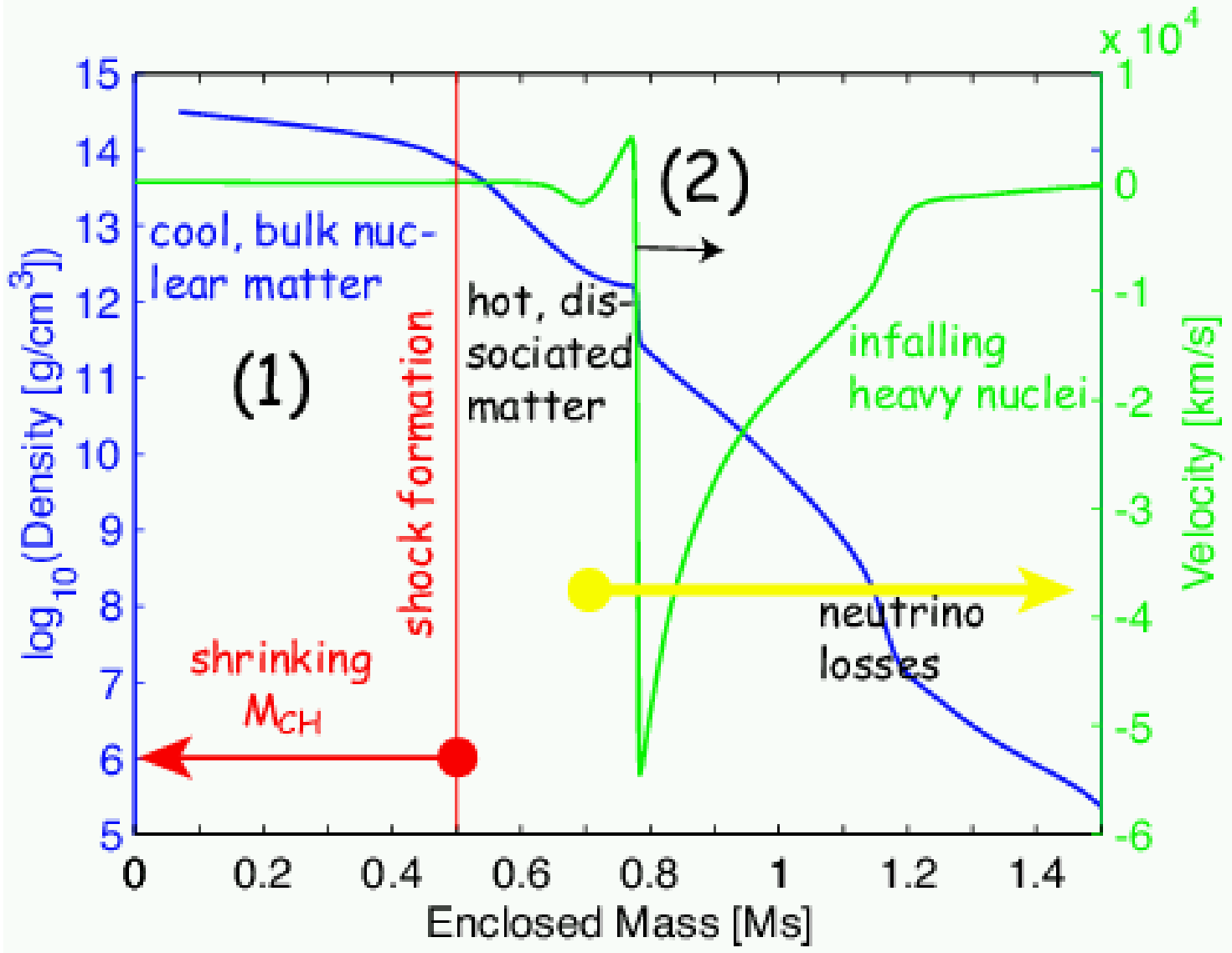}
\includegraphics[width=0.45\textwidth]{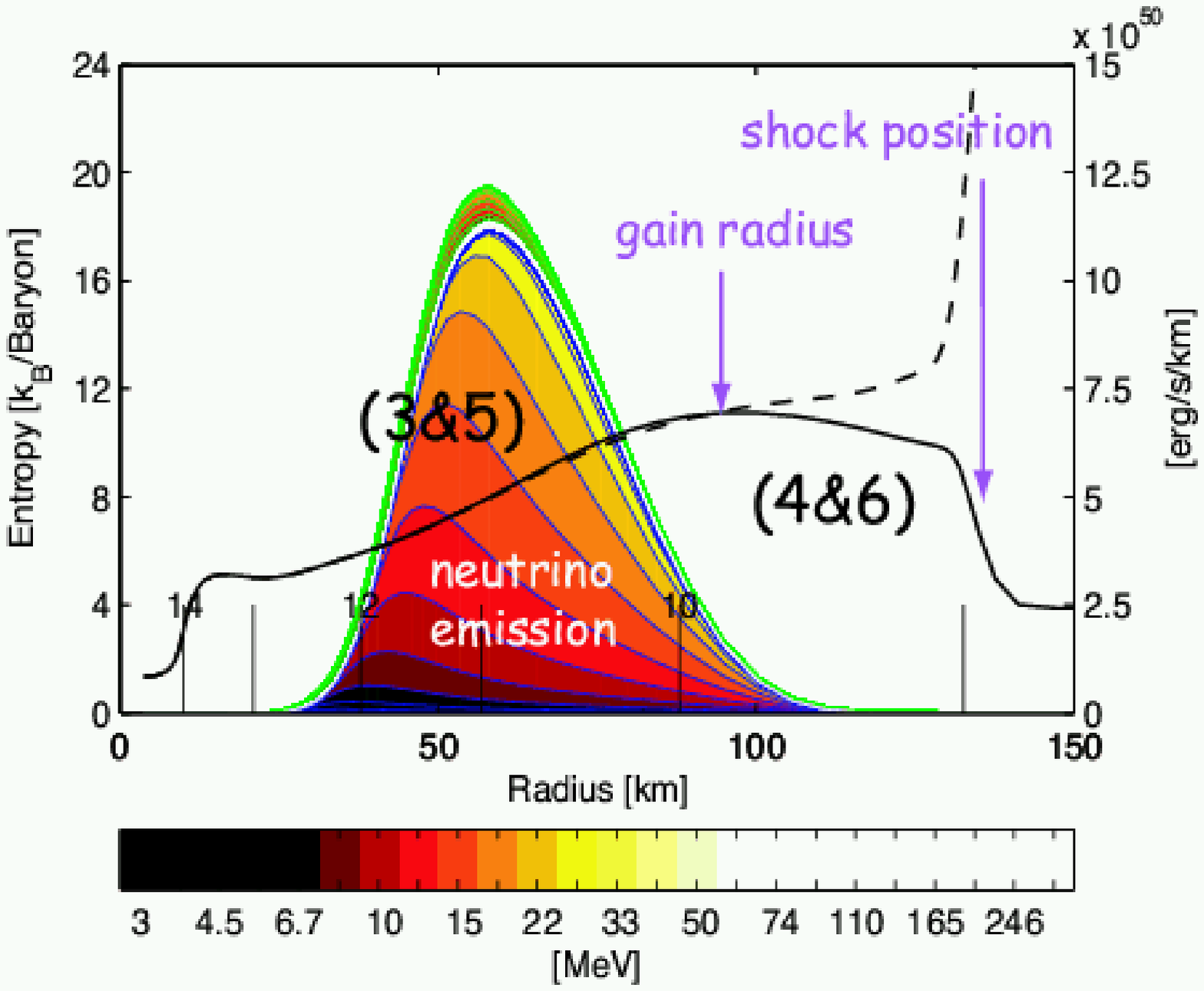}
\caption{Supernova models have to overcome at least one of six obstacles
before they can explode. These are: (1) electron capture during collapse,
(2) dissociation of nuclear matter by the shock, (3) neutrino cooling
of accreted material, (4) fast accretion velocities,
(5) the stability of the neutrinospheres, (6) technical challenges
with the neutrino transport. The graph on the left hand side shows
a density and velocity profile right after bounce, the graph on the
right hand side shows the entropy profile (solid line) at 100 ms
after bounce. The dashed line indicates the equilibrium entropy.
A more detailed description of the presentation is given in Ref.
\cite{Liebendoerfer_et_al_04}.}
\end{figure}

Spherically symmetric simulations ignore fluid instabilities
(for a recent discussion of convection in the heating region
and protoneutron star see e.g. Refs. \cite{Buras_et_al_03}
and \cite{Bruenn_Raley_Mezzacappa_04} respectively).
In spite of the agreement that fluid instabilities
favor explosions, the latter references suggest
that they might develop more slowly than hitherto assumed,
leading to failed supernova explosions in two-dimensional simulations
as well.
The computation time spent on hydrodynamics is likely to be negligible
in accurate supernova simulations in one or even two dimensions. Most
time is spent on energy-dependent neutrino transport. A systematic
improvement of the neutrino transport from one to two dimensions alone
required a substantial increase of computation time in yet incomplete
implementations \cite{Buras_et_al_03,Livne_et_al_04} and three-dimensional
neutrino transport has not yet been attempted with a reliable resolution
of the neutrino phase space. Following the three-dimensional simulations
of \cite{Fryer_Warren_02}, we try
to balance the computation time spent on hydrodynamics and neutrino
transport by maximizing the
degrees of freedom in the fluid dynamics in combination with approximations
in the neutrino transport. The effect of magnetic fields
on the dynamics of the nucleons in the hot mantle has not yet been
studied in three-dimensional numerical simulations with neutrino transport
approximations.

\section{An exploratory three-dimensional simulation with magnetic fields}

A simple and fast three-dimensional magneto-hydrodynamics code
\cite{Pen_Arras_Wong_03}
provides the core of our simulations. It spans a central
region of \( (600 \quad \mbox{km})^3 \) with an equidistant resolution of \( 1 \)
km in Cartesian coordinates. This covers the hot mantle and part of the
infalling layers. The code has received a new parallelization with MPI for
cubic domain decomposition that minimizes the resources occupied on distributed
memory machines by a simple and efficient reuse of buffer zones during
the directional sweeps. The finite differencing is second
order accurate in time and space and handles discontinuities in the
conservation equations with a total variation diminishing scheme.
Furthermore, a specific choice of the finite differencing for the
update of the magnetic field conserves its divergence to machine precision.
A velocity decomposition in the spirit of \cite{Trac_Pen_04} has
been implemented such that the entropy equation is solved for a smooth
large-scale bulk velocity and the total energy equation for
small-scale velocity perturbations. We
embed the computational domain of the MHD code in spherically
symmetrically infalling outer layers that are evolved by an implicitly
finite differenced one-dimensional hydrodynamics code
\cite{Liebendoerfer_Rosswog_Thielemann_02}.

Our collapse simulations are launched from a \( 13 \) M\( _{\odot } \)
progenitor model \cite{Nomoto_Hashimoto_88}.
The Lattimer-Swesty equation of state \cite{Lattimer_Swesty_91}
is used. We imposed
a rotation with angular velocity \( \Omega =4 \) s\( ^{-1} \) along
the z-axis with a quadratic cutoff at \( 100 \) km radius. Along
the same axis, we added a homogeneous magnetic field of \( 10^{12} \)
Gauss. The deleptonization during collapse has been parameterized
in a simplistic way: An investigation of the spherically symmetric
model N13 \cite{Liebendoerfer_et_al_03} with Boltzmann neutrino
transport reveals that the electron fraction
during infall can roughly be approximated as a function of density
\( \rho  \). In our three-dimensional simulation, we update the electron
fraction with \( Y_{e}(x,y,z)=\min \left[ Y_{e}(x,y,z),Y_{e}^{N13}\left( \rho (x,y,z)\right) \right]  \),
where the function \( Y_{e}^{N13}(\rho ) \) has been read out of
model N13 at the time of core-bounce. The effectiveness
of this simple parameterization for this exploratory simulation is
demonstrated in Fig.~2.

\begin{figure}[htb]
\begin{minipage}[t]{75mm}
\includegraphics[width=\textwidth]{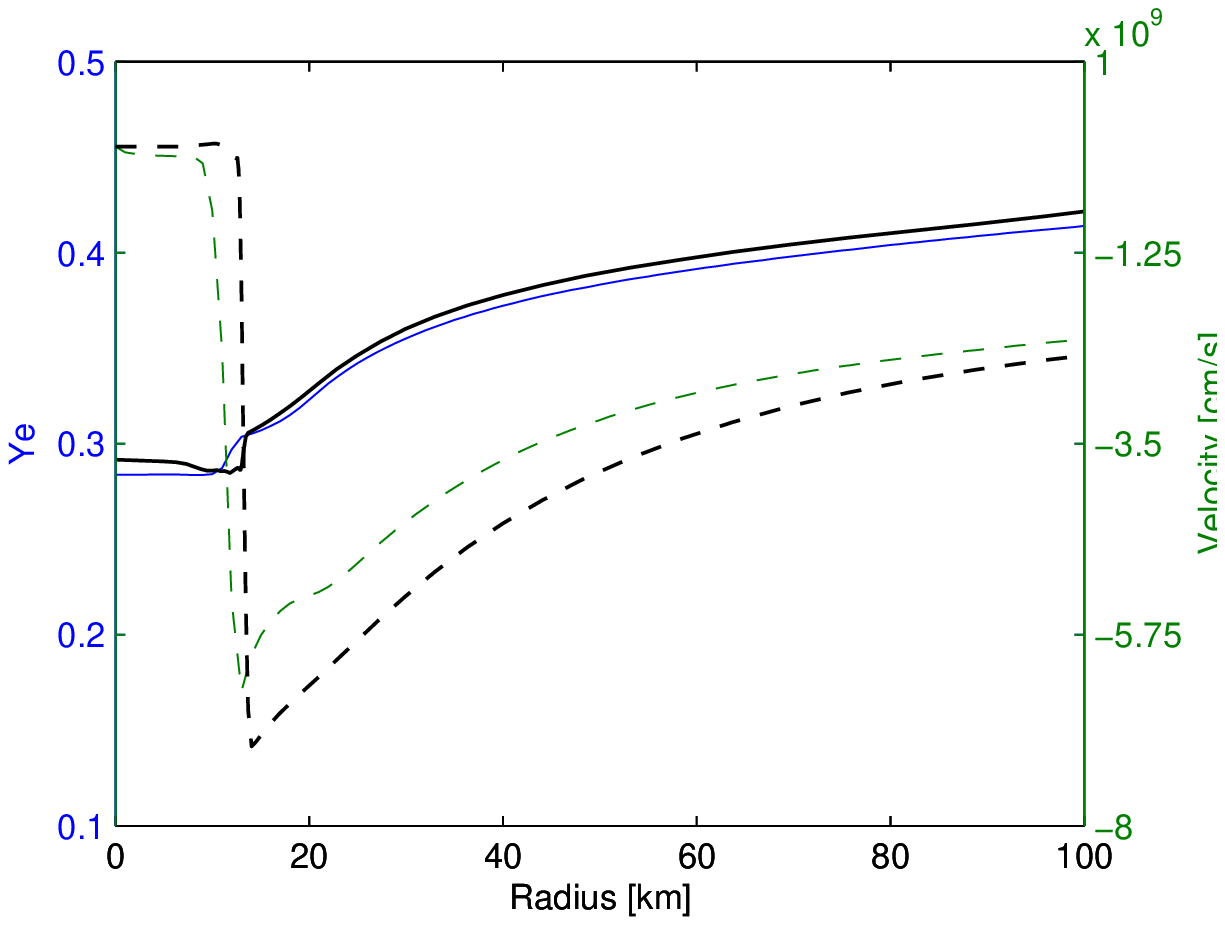}
\caption{Electron fraction (solid) and velocity (dashed)
profiles are compared between a spherically symmetric simulation
with accurate neutrino transport (thick lines) and the spherically averaged
quantities in the 3D MHD simulation (thin lines) at bounce.}
\end{minipage}
\hspace{\fill}
\begin{minipage}[t]{80mm}
\includegraphics[width=\textwidth]{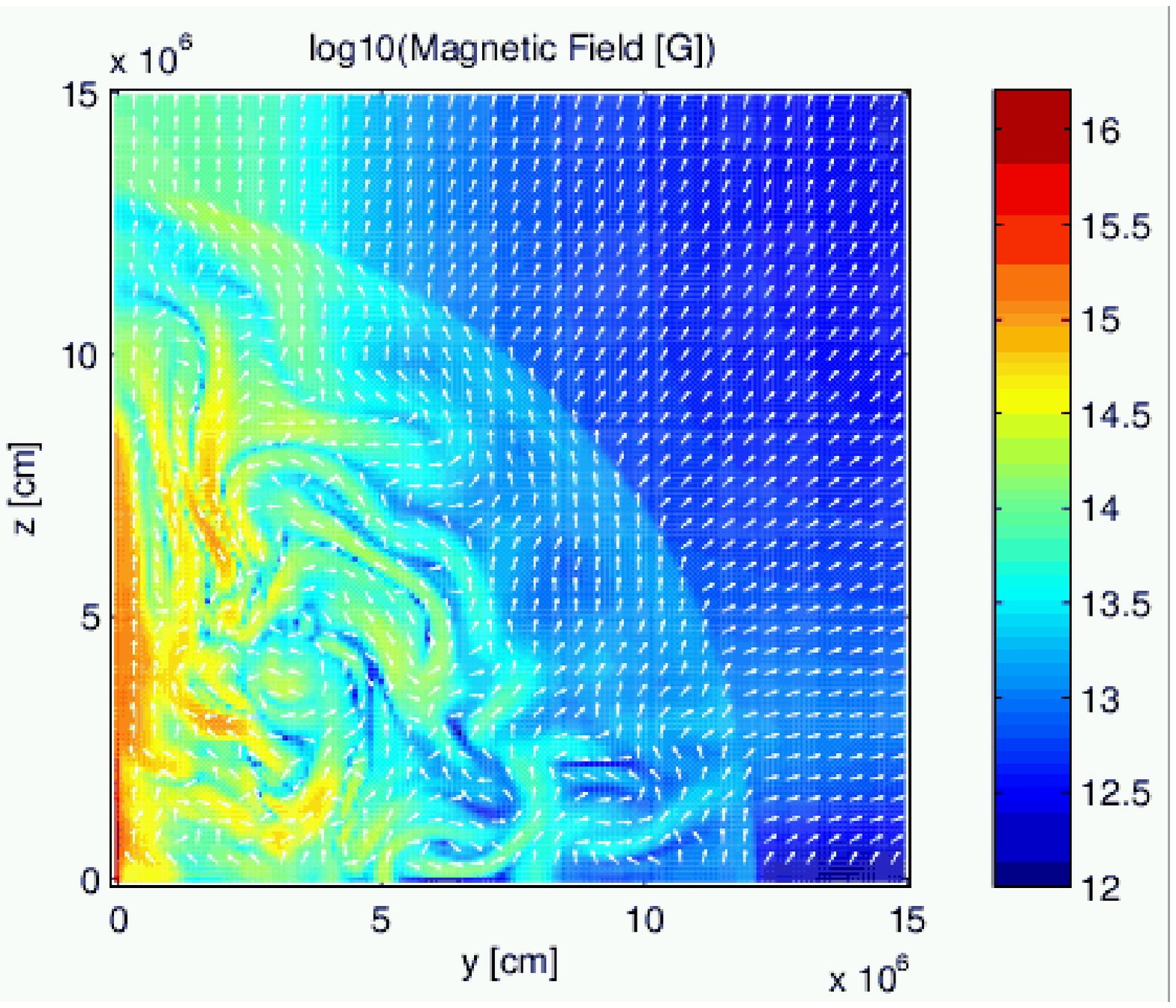}
\caption{The gray scale in the graph represents the logarithmic
amplitude of the entangled magnetic field in the y-z plane
at \( 15 \) ms after
bounce. The arrows indicate the directions of the field lines.
No symmetry is imposed in this three-dimensional simulation.}
\end{minipage}
\end{figure}

Due to the rotation, the polar infall
velocities are slightly larger than the equatorial infall velocities.
When the central density reaches \( 10^{11} \) gcm\( ^{-3} \)
the magnetic field lines become visibly distorted.
Due to the centrifugal forces, the projection of the velocities onto
the plane orthogonal to the initial magnetic field
is largest at about \( 100 \) km above and below the gravitational center.
These are
then the locations where the magnetic field lines condense most rapidly,
bending slightly outward around the center.
With ongoing collapse, this effect shifts to smaller radii and becomes
more pronounced.
At bounce, the magnetic field exceeds \( 10^{15} \) Gauss in these hot spots
located \( \sim 10 \) km above and below the center. The field lines run
along double cones aligned with the z-axis, except for the small deviation
that circumvents the center. In the early shock expansion until \( 5 \)
ms after bounce, the shock front is almost spherically symmetric.
Afterward, the simulation becomes unrealistic, because the dynamically
important neutrino burst is completely ignored. Behind the expanding
accretion front, entropy variations due to variations in the shock
strength induce fluid instabilities that entangle the magnetic field
lines. Fig.~3 shows an example snapshot at \( 15 \) ms after bounce.

The simulations have been performed on \( 64 \) processors of the
\( 528 \) processor McKenzie cluster at CITA. They required a wall clock
time of \( 150 \) hours. We demonstrated that three-dimensional magnetohydrodynamics
simulations in the supernova context are feasible at a comfortable
spatial resolution and that one-dimensional hydrodynamics is likely
to find a natural successor in future investigations of stellar core
collapse. As the neutrino interactions in the postbounce phase are
crucial and very sensitive to the neutrino energies, we plan to continue
with a multi-group neutrino light-bulb simulation to
study on a more realistic numerical foundation interesting
suggested effects of magnetic fields in the hot mantle (see e.g. Refs.
\cite{Thompson_00,Akiyama_et_al_03,Thompson_Quataert_Burrows_04,Kotake_et_al_04}).

\end{document}